# Decryption of Messages from Extraterrestrial Intelligence Using the Power of Social Media – The SETI Decrypt Challenge

René Heller

Max Planck Institute for Solar System Research, Justus-von-Liebig-Weg 3, 37077 Göttingen, Germany; heller@mps.mpg.de

**Abstract**

With the advent of modern astronomy, humans might now have acquired the technological and intellectual requirements to communicate with other intelligent beings beyond the solar system, if they exist. Radio signals have been identified as a means for interstellar communication about 60 years ago. And the Square Kilometer Array will be capable of detecting extrasolar radio sources analogous to terrestrial high-power radars out to several tens of light years. The ultimate question is: will we be able to understand the message or, vice versa, if we submit a message to extraterrestrial intelligence first, how can we make sure that they will understand us? Here I report on the largest blind experiment of a pretend radio message received on Earth from beyond the solar system. I posted a sequence of about two million binary digits ('0' and '1') to the social media that encoded a configuration frame, two slides with mathematical content, and four images along with spatial and temporal information about their contents. Six questions were asked that would need to be answered to document the successful decryption of the message. Within a month after the posting, over 300 replies were received in total, including comments and requests for hints, 66 of which contained the correct solutions. About half of the solutions were derived fully independently, the other half profited from public online discussions and spoilers. This experiment demonstrates the power of the world wide web to help interpreting possible future messages from extraterrestrial intelligence and to test the decryptability of our own deliberate interstellar messages.

## 1. Introduction

The identification of radio waves as a possible means for interstellar communication between distant intelligent beings (Cocconi and Morrison, 1959) sparked the modern search for extraterrestrial intelligence (SETI). Early attempts of Project Ozma at the National Radio Astronomy Observatory at Green Bank (Drake, 1961) were followed by more than a hundred independent SETI projects within the next 40 years (Tarter, 2001) and even led to the construction of dedicated SETI infrastructure such as the Allen telescope array (Tarter et al., 2011). Modern SETI has culminated in the privately funded SETI by the Breakthrough Listen Initiative[1], which searches both in the radio (Isaacson et al., 2017; Gajjar et al. 2017; Enriquez et al. 2017a; Enriquez et al. 2017b) and optical (Tellis & Marcy 2017) regimes of the electromagnetic spectrum. In the next decade, commensal SETI observations will also be made with the Square Kilometer Array (Siemion et al., 2015), the most sensitive radio telescope ever built. Hence, even though it is impossible to reliably predict the chances of establishing interstellar contact within our generation, addressing the question of whether humans would actually be able to understand the content of an interstellar message is timely.

Beyond the question of which methods to use for communication, a key challenge for first contact interstellar communication is in the choice of the language, known as the incommensurability problem (Vakoch, 1999). How could an intelligent transmitter make their message interpretable by a distant intelligent receiver if they don't even know whether the other party even has a clue about symbols as a means of information delivery? Proposed solutions include submissions of constructed languages and logical symbols (Freudenthal, 1960; Busch and Reddick, 2010) or a combination of basic mathematical symbolism (e.g. binary digits representing numbers) with images as used for the Arecibo message (Staff at the National Astronomy and Ionosphere Center, 1975) and aboard the Pioneer and Voyager spacecrafts (Sagan et al., 1972).

[1] https://breakthroughinitiatives.org/initiative/1

A realistic test of an interstellar message on Earth would invoke two technological civilizations that developed their skills of modern astronomy independently. Although different human cultures have now developed astronomical and even astronautical expertise, their intellectual independence is not given on Earth due to the modern global communication network. Hence, we can hardly simulate such a scenario and a truly blind test is impossible.

Nevertheless, one such limited approach is in submitting a pretend SETI message to colleagues, who are not privy to the message design or content, and then let them try to decrypt the message. This has been done before (Busch and Reddick, 2010) and it goes back to early attempts by Drake in 1962, who tested the decryptability of an early version of the Arecibo Message (a sequence of binary digits representing an image) by sharing it with his nine fellow participants of the 1961 SETI conference at Green Bank (see Figure 5 by Vakoch, 1999). According to Drake and Sobel (1992), Drake received a single answer to his message, which was a sequence of binaries from Bernard Oliver and which, if translated into a black/white pixel map, showed an olive in a Martini glass. While the receiver of the initial message (Oliver) obviously had recovered the image nature of the original binary sequence, its content was hardly interpretable even by an educated human. Intriguingly, about a year later, Drake received a letter from an electronics engineer, who had seen Drake's message in a magazine for amateur code crackers, and who had then indeed decrypted Drake's message successfully.

All these previous attempts of testing the decryptability of SETI messages suffer from the fact that the seemingly blind receiver has, in fact, a cultural and educational background that is usually similar to the one of the transmitter. Moreover, the test audience of these previously conceptualized SETI messages was limited to just a few people. This limitation can now be overcome through the internet or world wide web (WWW).

This paper reports on what can be considered the first open, transparent, live collaboration of scientists and non-scientists on the decryption of a pretend, incoming interstellar message. Our experiment can be seen as a hypothetical release of a newly discovered SETI message in accordance with the principles proposed by the International Academy of Astronautics (IAA Position Paper, 1989). Here we discuss a novel and hitherto unexplored method to test the decryptability of SETI message by making it available to all humans with an internet access, most of which have an educational background that is very distinct from the one of the sender, i.e. the author of this paper. The aims of this experiment were twofold. First, we wanted to test the ability of the social media to help decrypting potential future messages from extraterrestrial intelligence. Second, we wanted to learn whether people, who are untrained in SETI decryption, would be able to read the message. This second aspect is maybe the most realistic blind test of a SETI message conducted as of today, and it might be helpful for humans to encrypt their own messages with the aim to make them as easily understandable as possible.

## 2. The SETI Decrypt Challenge

On Tu, 26 April 2016, I submitted a fake SETI message to the social media Twitter[2] and Facebook and called out the SETI Decrypt Challenge. The original call is given in the appendix of this manuscript. The deadline for submissions was 3 June 2016.

The message can be considered an interstellar greeting card from a pretend alien sender in the sense that it broadcasts merely the basic facts about their physical environment. Its volume of 1,902,341 characters (or Boolean values, to be precise) is about a thousand times larger than the 1,679 binary digits of the Arecibo message (Staff at the National Astronomy and Ionosphere Center, 1975), but still much more simple than the lingua cosmica ('Lincos') mathematical language developed by Freudenthal (1960) or its modern version CosmicOS.[3]

### *2.1 The frequency and general content of the message*

The fictional detection scenario of this message assumes a series of radio pulses from a fixed, unresolved source about 50 light years from Earth, which are received in a very narrow band around an electromagnetic frequency

[2] https://twitter.com/DrReneHeller/status/724935476327624704

[3] http://cosmicos.github.io

```
(line 1)
111111111111111111111111111111111111111111111111111111111111111111111111111111111111111111
111111111111111111111111111111111111111111111111111111111111111111111111111111111111111111
111111111111111111111111111111111111111111111111111111111111111111111111111111111111111111
11111111111111111111111111111111111111111111111111111111111111111111111111111111111111111

...
(line 2)
000000000000000000000000000000000000000000000000000000000000000000000000000000000000000000
000000000000000000000000000000000000000000000000000000000000000000000000000000000000000000
000000000000000000000000000000000000000000000000000000000000000000000000000000000000000000
00000000000000000000000000000000000000000000000000000000000000000000000000000000000000001

...
(line 3108, or line 80 on page 5)
000000000000000000000000000000000000000000000000000000000000000000000000000000000000000000
010101000000000000000000000001000101010101010101010101010101010101010101010101010101000000
000101010001000100000000000001010101000100000000000000000000000000000000000000000000000000
00000000000000000000000000000000000000000000000000000000000000000000000000000000000000000
```

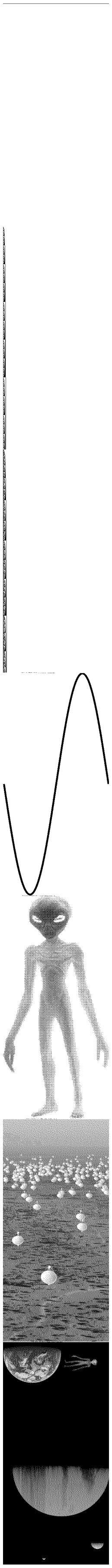

**FIG. 1** *Top:* Sequence of binary digits that constitute the pretend SETI message. Three excerpts are shown. The entire message contains 7 × 359 × 757 = 1,902,341 bits. The lines denoted on top of each excerpt refer to an arrangement of 359 pixels (width) times 757 pixels (height) of this string of binary digits. *Right:* Arrangement of the message on a 359 × 757 pixel map, where '1' is converted into a black pixel and '0' is converted into a white pixel. This conversion is arbitrary, the negative of this image contains the same information.

of 452.12919 MHz.

Regarding the transmission frequency, the radio regime of the electromagnetic spectrum has been identified as an optimal range of wavelengths because radio waves are both hardly absorbed in planetary atmospheres or by the interstellar medium and they do not demand high power at the source to be detectable against the stellar radio emission (Cocconi and Morrison, 1959). Some have proposed the 'water hole' between the hyperfine transition of neutral atomic hydrogen at a frequency of 1,420 MHz (a wavelength $\lambda$ = 21.1 cm) and the principal hydroxyl microwave transition frequency at 1,667 MHz ($\lambda$ = 18.0 cm) (Oliver and Billingham, 1971) as an optimal range of channels. The attractiveness of the 21.1 cm line, however, is decreased by the noise in the galactic plane. Other natural frequencies can be constructed from the fundamental physical constants and considerations of the cosmic microwave background (Drake and Sagan, 1973). For my pretend message, I used a frequency of 452.12919 MHz, corresponding to $\lambda = \pi \times 21.1$ cm = 66.3 cm, a frequency that has also been mentioned in the 1997 science fiction movie 'Contact'.

As an aside, a possible Doppler shift of the fake source can be neglected. Even if this message came from a planetary system around a high-velocity star with a relative motion to the Sun of 100 km $s^{-1}$, or about 1/3,000 the speed of light, its relativistic Doppler effect would be a mere $3 \times 10^{-4}$, that is, 0.2 mm at a wavelength of about 66 cm or roughly 0.15 MHz at a frequency of about 452 MHz. Even if the message were detected at a frequency shifted by as much as 0.15 MHz, this should still allow the receiver to identify this frequency as the product of the two said constants. Another contribution to the relativistic Doppler shift of the message could come from the orbital motion of the sender around its host star. If it were located on an Earth-like planet around a Sun-like star, e.g., then for a circular orbit we should observe a sinusoidal component with a period of roughly one year and with an amplitude of some 30 km $s^{-1}$ or about 1/10,000 the speed of light.

It has been noted that SETI messages using constructed language have a higher information density (Busch and Reddick, 2010). On the other hand, humans proverbially agree that a picture paints a thousand words. Hence, if extraterrestrial civilizations have at least some kind of a sensuous nature, then they might be susceptible to the power of images as well and they might want to use them to transmit their own impressions from what they would call home. Sending a message that contains only math or symbolic language might give the receiver an idea of the syntax of language but it would hardly convey what the symbols mean. Hence, pictures will be necessary to link symbols with objects. Moreover, spoken and written language came up rather late in the evolution of the hominids, which is why pictures will also be necessary to convey how we think.

**TABLE 1.** Examples of little-endian binary code translated into decimal numbers.

| Binary Notation | Meaning | Decimal number |
|---|---|---|
| 000 | $0 \times 2^0 + 0 \times 2^1 + 0 \times 2^2$ | 0 |
| 100 | $1 \times 2^0 + 0 \times 2^1 + 0 \times 2^2$ | 1 |
| 010 | $0 \times 2^0 + 1 \times 2^1 + 0 \times 2^2$ | 2 |
| 110 | $1 \times 2^0 + 1 \times 2^1 + 0 \times 2^2$ | 3 |
| 001 | $0 \times 2^0 + 0 \times 2^1 + 1 \times 2^2$ | 4 |
| 111 | $1 \times 2^0 + 1 \times 2^1 + 1 \times 2^2$ | 7 |
| 1010111001101 | $2^0 + 2^2 + 2^4 + 2^5 + 2^6 + 2^9 + 2^{10} + 2^{12}$ | 5749 |

The pretend SETI message of this experiment has thus been tailored in a way to combine pictures with a constructed language.

I decided that the hypothetical senders of the message wanted to tell us something about themselves (a selfie of the alien body was chosen as an appropriate image to be decrypted in the social media), about their technologies, and about their planetary system.

### *2.2 Spatiotemporal, universal yardsticks*

The challenge in encoding the spatial and temporal information that would come along with the images was in choosing unequivocal, universally understandable yard sticks. I chose the modified versions of the Planck length ($l_P$) and Planck time ($t_P$). These quantities are part of a metric system that is based on natural constants as base units only, and they can be derived from the natural constants $\hbar = h/2\pi = 1.054 \times 10^{-34}$ Js (the Planck constant), $c$ = 299,792 458 m s$^{-1}$ (the speed of light), and $G = 6.674 \times 10^{-11}$ m$^3$ kg$^{-1}$ s$^{-2}$ (the gravitational constant) as per $l_P = (G\,\hbar/c^3)^{1/2} = 1.616 \times 10^{-35}$ m and $t_P = l_P/c = 5.391 \times 10^{-44}$ s.

For the message, however, I arbitrarily chose a modified Planck length $l_P' = (G\,h/c^3)^{1/2} = 4.05 \times 10^{-35}$ m and the modified Planck time $t_P' = l_P'/c = 1.3512 \times 10^{-43}$ s to simulate a genuine normalization of the Planck length by the transmitting party. Then the wavelength of the message (66.3cm) equals $1.6368 \times 10^{34}$ times $l_P'$ and the time that the message traveled from the sender to Earth equals $1.1677 \times 10^{52}$ s times $t_P'$.

### *2.3 The message*

The message contains 7 × 359 × 757 = 1,902,341 bits, where the factors 7, 359, and 757 are prime numbers. When properly arranged, e.g. simply using a text editor, the string of binary digits can be easily displayed as a sequence of seven pages, each of which has a width of 359 pixels and a height of 757 pixels. The first page contains a 359 pixels by 757 pixels configuration frame that defines the size of the pages (see Figure 1). Page two presents a simple count-up from 0 to 756 using little-endian binary, where the least significant digit is printed first. Table 1 lists a few examples of how to convert from little-endian binaries into decimals. The third page presents a list of the first 757 prime numbers, the largest of which is 5749. Pages two (simple counting) and three (prime numbers) were meant to establish that we are talking binary and that we use numbers to describe and interpret the images.

Page four contains an image, which can be constructed from the binary code converting zeros into white pixels and ones into black pixels or vice versa. This image is an illustration of the function -sin($x$), where the negative algebraic sign is only chosen to prevent the sine curve from blending into the image header that consists of two lines. The first of these two header lines is a number close to $1.6368 \times 10^{34}$, the second line contains a number close to $1.1677 \times 10^{52}$. The first line can be identified as the spatial dimension since it denotes the wavelength in units of the modified Planck length. To be clear: the receiver (us) would know the number in the header, the wavelength that the message was transmitted on, and the modified Planck length. The challenge would be in putting them into the said context, that is, to seize that the binary-encoded number equals $l_P' \times \lambda$. The second line can be identified as the time component: divide the modified Planck length by the speed of light and one will obtain a value very close to the modified Planck time. Alternatively, we can multiply our now identified temporal yardstick, $t_P$', with that huge number given in the header to find that the result of this operation is 50 yr. As we know that the message comes from a star 50 ly away, our alien friends are telling us that they know their distance to us as well. The wave period could have provided a more plausible yard stick to define the temporal scale rather than the time, which the message had been traveling. Hence, our interstellar counterparts seem to have the intention to let us know that the message is really a 'private' message for us, the (potential) inhabitants of the solar system. Whether they have indeed found us and detected our interstellar beacons and signs of intelligence, however, we don't know. Nevertheless, we actually have some additional information coming with the message that is beyond the pure images and numbers.

On page number five, we see an image of one specimen of our interstellar counterparts. The header of this image contains binary representations of the numbers $6.048 \times 10^{34}$ and $4.2038 \times 10^{52}$, which translate into a length of 2.45 m and a time of 180 yr. These numbers can be interpreted as the height and age of the alien body, potentially as their typical body size and lifetime. The sixth page shows an image of the device used by the sender to submit their transmission. The image is actually an excerpt of an artist's rendering of the 5 km diameter central core of the SKA[4]. Page seven shows four astronomical objects, one of which is being hovered by a graphic of the specimen. Its header can be converted into 0.26 astronomical units (AU), or almost exactly 100 times the Earth-Moon distance, and 6 billion years. The latter information is suggestive of the lifetime of the aliens' host star system, whereas the spatial information really was a double-blind test. I did have a specific astronomical scale to encode in mind (100 times the Earth-Moon distance), but I did not know the spatial context of the submitting party. Instead, I was interested in the ideas of the participants of the decrypt challenge as to what this number could actually signify. That said, the incidence of light on the astronomical bodies shown on the image suggests that the light source, that is to say the star, is far off the image margins.

An audio version of the message is available at https://soundcloud.com/user-165653195/seti-decrypt.

I asked the following six questions, which could be answered if the message was successfully decrypted. The answers are given in brackets.

1. What is the typical body height of our interstellar counterparts?
(2.45 m)

2. What is their typical lifetime?
(180 yr)

3. What is the scale of the devices they used to submit their message?
(100 km)

4. Since when have they been communicating interstellar?
(10,000 yr)

5. What kind of object do they live on?
(a moon or planet, 0.26 AU = 100 times the Earth-Moon distance)

6. How old is their stellar system?
(6 Gyr)

### *2.4 Automated preparation of the message*

To avoid the need of manually typing a chain of two million zeros or ones, I wrote a python script to do the job in less than one second.[5] As inputs, it uses the images shown on pages four to seven of the message (see Fig. 1) in a portable bit map (PBM) format, then writes the corresponding bits into the message, and finally returns both a PBM image and a text (TXT) file of the entire message. The natural constants ($c$, $G$, $h$) and the wavelength of the message are defined in the first few lines of the code, followed by the reading of the input files and their conversion into 757 strings of 359 bits to give one page. Each header of a page, i.e. the little-endian binary code translation of the tempo-spatial yardstick, is calculated and written on-the-fly for each page.

[4] Available at https://commons.wikimedia.org/wiki/File:SKA_overview.jpg. Contributed by Wikipedia user Skaoutreach. Licensed under the Creative Commons Attribution-Share Alike 3.0 Unported license.

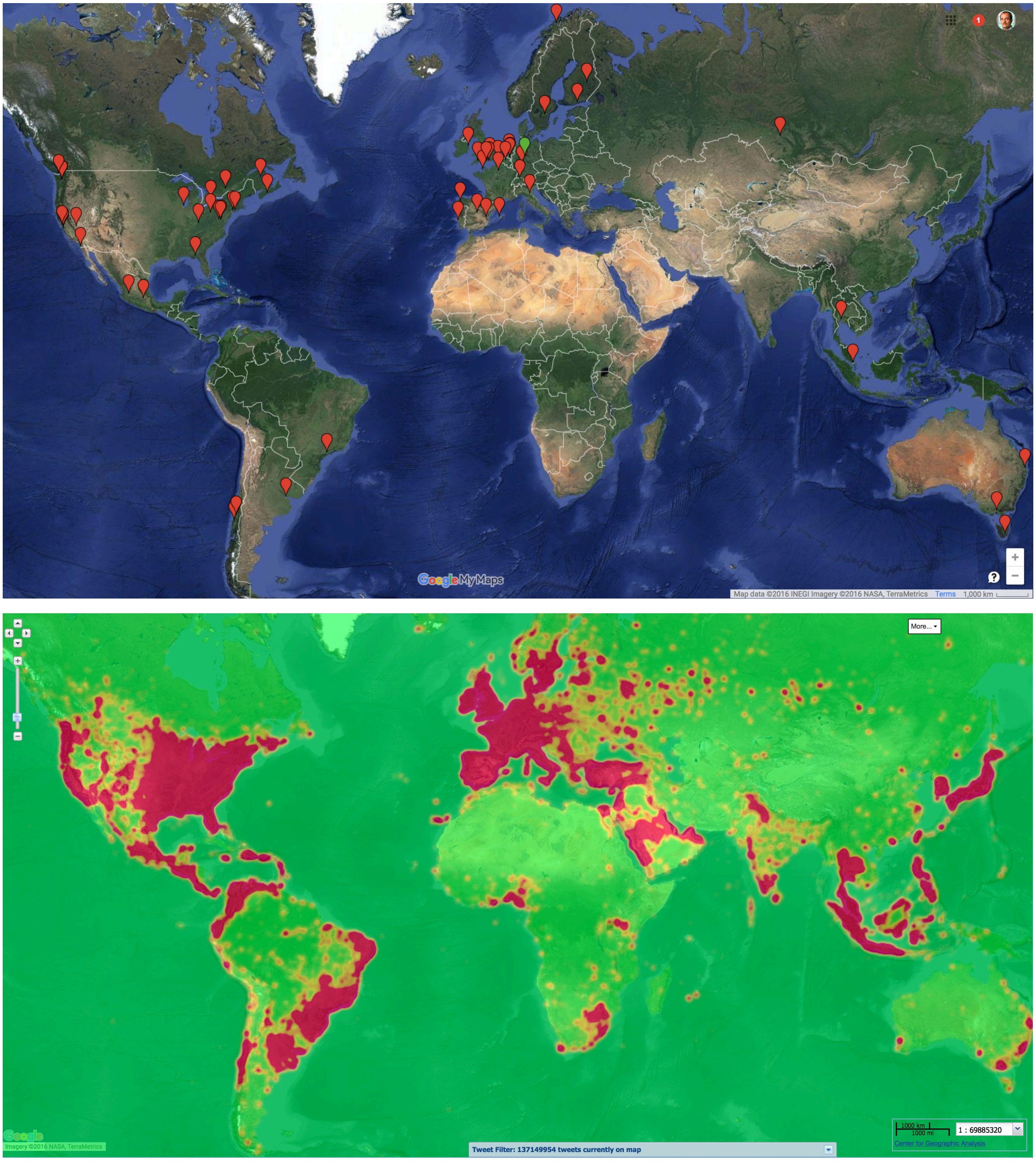

**FIG. 2**: *Top*: Geographical distribution of the 66 successful teams of the SETI Decrypt Challenge. *Bottom*: Twitter heat map based on 130,444,624 tweets analyzed between 3 and 17 December 2013. Image credit: Map data ©2016 INEGI Imagery ©2016 NASA, TerraMetrics; Center for Geographic Analysis (https://worldmap.harvard.edu/tweetmap).

[5] The code is freely available upon request via e-mail to the author.

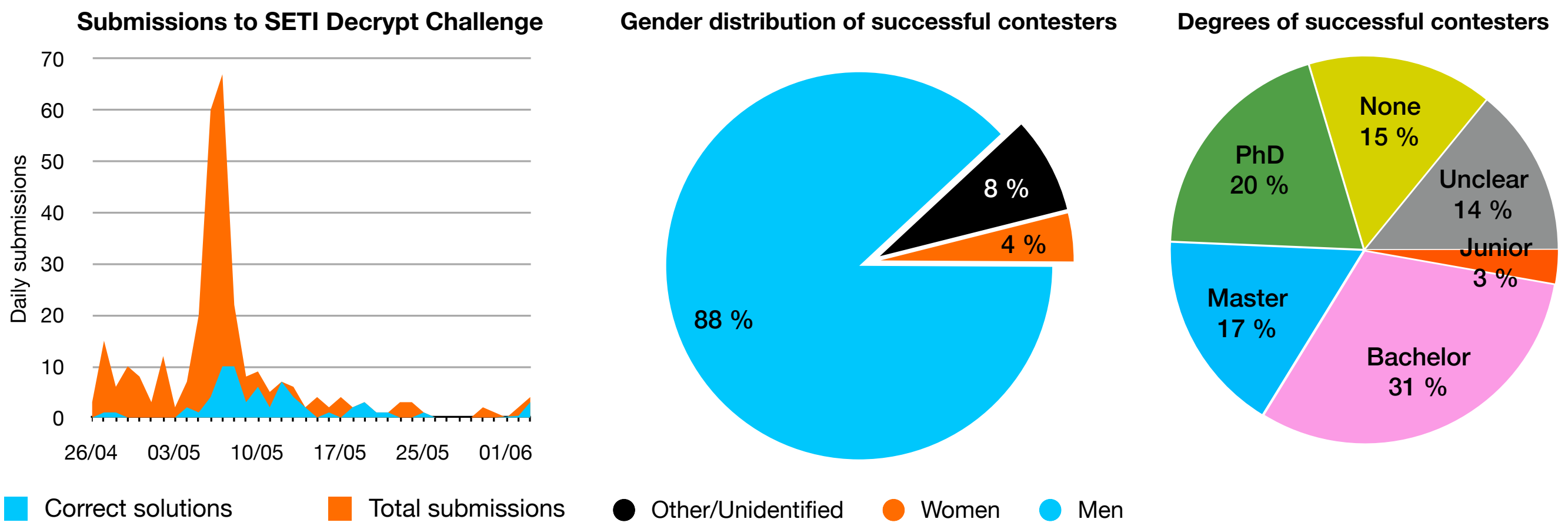

**FIG. 3**: Analysis of the activity and the gender/academic composition of the SETI Decrypt contesters. *Left*: Correspondence with contesters. The peak on 07/05/2016 might be correlated with online articles of the Daily Mail or the Huffington Post, or with a spoiler posted on reddit.com in the evening hours (EST) of the same day. In the first few weeks, the fraction of correct solutions was small, while most submissions after the peak were correct. *Center*: A mere 4% of the successful contesters were identified as women and 8% were unclear or neither women or men, while 88% were identified as men. *Right*: Most of the decrypters had a bachelor degree (31%) or a PhD (20%), mostly in natural sciences. Two of the successful decrypters (or 3%) were junior school students. The entire sample consists of 71 individuals, spread over 66 contesters including teams.

## 3. Results

### *3.1 Dissemination of the message and response activity*

The original message was posted on Twitter and Facebook on 26. April 2016. The first solutions were submitted to the author the next day and the corresponding twitter activity gained momentum over the next week, when the original twitter posting (the tweet) was retweeted or favorited several times a day. I also received several replies via e-mail per day over the first ~7 days after the posting. The website of The Planetary Habitability Laboratory of the University of Puerto Rico at Arecibo featured the call on 27 April.[6] On 29 April, the tweet was also posted on reddit.com[7], where an discussion generated 56 comments within three days. Later, the Daily Mail[8] (5 May) and the Huffington Post[9] (6 May) reported on the challenge, which could be the reason for a dramatic increase of submissions to the challenge over the following two days.

Sixty-six correct submissions, mostly via e-mail to the author, were received between 26 April and 3 June. People were allowed to build teams. Seventy-one people were involved in the 66 correct submissions. Figure 2 shows the geographic distribution of the successful challengers, where the position of a red marker corresponds to the contemporary location (upper panel). The density of those markers naturally correlates with the twitter heat map (lower panel), because this is the social medium where the original call had been posted. Most of the successful challengers came from Western Europe and the North American east coast. Africa was completely absent, as were China, India, and most of South America, Russia, and Australia. This might, at least partly, be due to the fact that the call was posted in English.

---

[6] http://phl.upr.edu/library/notes/SETIChallenge

[7] https://www.reddit.com/r/space/comments/4gz75v/can_you_decrypt_this_alien_message

[8] http://www.dailymail.co.uk/sciencetech/article-3575382/Can-solve-alien-code-Physicist-reveals-tricky-number-challenge-humans-crack-ET-s-message.html

[9] http://www.huffingtonpost.co.uk/entry/astrophysicist-ren%C3%A9-heller-sets-alien-code-binary-code-test-could-you-crack-it_uk_572c573be4b0e6da49a5fe24

### *3.2 Distribution of gender and academic degrees*

Each of the successful decrypters was asked to kindly provide their current location of residence and their highest academic degree. They were not asked for their gender, but this information would be derived from the contesters‘ clear names or twitter account profile pictures (or both if available). All of this information is available to the author on a person-by-person base but will be anonymized for this report.

The left panel of Figure 3 shows the number of daily submissions. Orange denotes the total number of submissions via e-mail, twitter, and facebook per day. The orange area corresponds to 305 submissions in total. This includes failed and correct solutions, questions, bids for hints, responses to the author's follow-up requests, reports of unrelated contacts with aliens, and insults. The peak of 66 submissions on 7 May could be related to the above-mentioned online articles of the Daily Mail or the Huffington Post or to a spoiler on reddit.com that I was informed about by one of the contesters. Light blue encodes the number of correct submissions per day. Interestingly, while only few of the submissions prior to 7 May were in fact correct solutions, most of the submissions that were received after the posting of the spoiler were correct.

The central panel of Figure 3 shows the gender distribution of the successful contesters: 88% were men, 8% were either not male and not female or unidentified, and 4% were women.

The right panel of Figure 3 displays the distribution of the highest academic degrees of the contesters, almost half of which had either a bachelor (31%) or a PhD (20%), mostly in natural sciences. The other half had a master (17%), no academic degree (15%) or were unclear in this regard (14%). We also saw two juniors (one 10th grade and one 11th grade high school student) solving the SETI Decrypt Challenge, making up about 3% of the contesters.

The programming languages used by the competitors include Matlab, R, python, C, C#, bash, and Javascript.

## 4. Discussion

The definition of the frame size on page one was recognized by most of the people who replied to the call. Many people were also able to put the wavelength at which the message had been received (66 cm) in context with the binary number that was intended to contain the spatial information, although the wavelength had not explicitly been mentioned in the call, but rather the frequency of the message (452.12919 MHz).

Some participants noted that the spatial and temporal yardsticks were very similar to the Planck numbers, but off by a factor of about 2.5. In fact, they were off by a factor of $(2\pi)^{1/2} = 2.5066$. Interestingly, as most of the contesters were able to synchronize the spatial measure with the wavelength and the temporal measure with the time that the message had been traveling (50 yr), they were nevertheless able to convert the numbers in the headers of pages four to seven into the correct secular scales and units. Many people interpreted the temporal number given with the picture of the wave as the wave period. It remains to be shown, e.g. in an improved, more realistic decryption experiment, that the quantities in the header would naturally be interpreted as the correct physical attributes of the subject shown if there would be no questions biasing accompanying the message.

One participant noted that the pixel width of the image (359) is the largest prime number that is still smaller than (and very close to) the number of degrees in a circle. This was pure coincidence because the hypothetical senders of the message were not expected to divide circles by an arbitrary number of 360 as we use to do.

One of the successful decrypters came up with a neat way of helping others who were still struggling with the message to test if their solution is correct: 'Alien height (m) x life span (yrs) x antenna size (m) x age of solar system (B yrs) / how long been transmitting (yrs) = 26,460.'

This experiment assumes that most aspects of the hypothetical message have an unambiguous interpretation or solution. In a real case scenario, however, multiple interpretations of the whole message could be equally plausible. In fact, the message might be ambiguous by design. While the proposed solutions filed by the contesters of the SETI Decrypt Challenge were either rejected as 'wrong' or accepted as 'correct' by the author, a real message might not even foresee a single correct interpretation.

## 5. Conclusions

Although 66 teams, including 71 individuals, submitted their correct solutions mostly by e-mail, it can be assumed that these solutions have not been derived entirely independently. After about 30 submissions (9 May 2016), it occurred to us that decrypters referred to hints and suggestions they had picked up in the internet, which might also be related to the online spoiler mentioned above. Although this fact distorts the results of this experiment to some extent, it also highlights the strong potential of collaboration on SETI message decryption in the WWW, possibly based on data to be retrieved by the Breakthrough Listen Initiative (Isaacson et al., 2017).

Beyond decryption, the results of the SETI Decrypt Challenge could be useful for the concerted effort to design a human-made message to the stars, be it using the Square Kilometer Array (Siemion et al., 2015) or in the framework of Breakthrough Message[10].

Regarding the content of man-made interstellar messages, it has been suggested earlier that humans could call out by sending Big Data, e. g. the entire content of the internet using powerful lasers (Shostak, 2015). We argue instead that interstellar exhibitionist behavior entails a vast amount of critical issues, such as divulgement of our military soft spots. There is knowledge that is meant for humans only. We propose that humans should rather start by sending a concise, well-thought, and balanced message just like it is human to show one's open palm or to shake hands first rather than to exchange the details of one's bank accounts or private photographs. Social media offer a forum for humanity to develop such a concerted message, e.g. by testing its decryptability.

Ultimately, as proposed and adopted by the SETI Committee of the International Academy of Astronautics in 1989, 'a confirmed detection of extraterrestrial intelligence should be disseminated promptly, openly, and widely through scientific channels and public media'. Consequently, any received SETI message would be released to the WWW soon upon detection. The SETI Decrypt Challenge experiment demonstrates that the WWW social media offer an active forum and a vast palette of expertise in different scientific areas to the decryption of SETI messages. Hence, their distribution in the social media would not only offer an efficient means of decryption but also offer an unprecedented opportunity to unite humans all over the globe in a common scientific and cultural effort.

## Appendix A.

This is a call for a fun scientific challenge.

Suppose a telescope on Earth receives a series of pulses from a fixed, unresolved source beyond the solar system. The source is a star about 50 light years from Earth. The pulses are in the form of short/long signals and they are received in a very narrow band around an electromagnetic frequency of 452.12919 MHz. A computer algorithm identifies the artificial nature of the pulses. It turns out the pulses carry a message. The pulses signify binary digits. Suppose further that you were, by whatsoever reason, put in charge of decrypting this message.

If you successfully decrypted the message, you should be able to answer the following questions:

1. What is the typical body height of our interstellar counterparts?
2. What is their typical lifetime?
3. What is the scale of the devices they used to submit their message?
4. Since when have they been communicating interstellar?
5. What kind of object do they live on?
6. How old is their stellar system?

These are the rules.

[10] https://breakthroughinitiatives.org/initiative/2

1. No restrictions on collaborations.
2. Open discussion (social networks etc.) of possible solutions strongly encouraged.
3. 3 hints to the solutions can be offered as per request.
4. Send your solutions to me via e-mail (heller@mps.mpg.de), twitter (@DrReneHeller) or facebook (DrReneHeller). Human-readable format and the format of the message are allowed.
5. On 3 June 2016, a list of the successful SETI crackers (in chronological order) will be released.

UPDATE 6 May 2016:
This call generated an e-mail storm on me. I kindly ask you for your understanding that I will restrict replies via e-mail to a minimum. Correct submissions will, of course, be acknowledged.

UPDATE 7 May 2016
E-mail traffic is still overwhelming. From now on, I will not be able to give additional hints and will only respond to correct (or very creative) solutions via e-mail in English or German.

These are the three hints mentioned in the rules.

1. The number of bits (0 or 1) is 1902341. This is a product o the prime numbers 7, 359, and 757.

2. The message is the black/white pixel map of an image.

3. The image shows 7 pictures or pages. As a sanity check, you will be able to recover the duration of the travel time (50 years) from page 4.

## Acknowledgements

This online SETI decryption experiment was inspired by the author's reading of the book ‘Is anyone out there? The scientific search for extraterrestrial intelligence’ by Frank Drake and Dava Sobel. The author is thankful to the referee report of an anonymous reviewer. This work was supported in part by the German space agency (Deutsches Zentrum für Luft- und Raumfahrt) under PLATO Data Center grant 50OO1501. This work has made use of NASA’s Astrophysics Data System Bibliographic Services.